\documentclass[]{article}
\usepackage[dvips]{graphicx}

\begin{document}

\title{Dominance of hole-boring radiation pressure acceleration regime with thin ribbon of ionized solid hydrogen}

\author{J Psikal$^{1, 2}$, M Matys$^1$}

\date{
  $^1$ Faculty of Nuclear Sciences and Physical Engineering, Czech Technical University in Prague, Czech Republic\\
  $^2$ ELI-Beamlines project, Institute of Physics, Czech Academy of Sciences, Czech Republic\\[2ex]
  \today  
}

\maketitle

\begin{abstract}
Laser-driven proton acceleration from novel cryogenic hydrogen target of the thickness of tens of microns irradiated by multiPW laser pulse is investigated here for relevant laser parameters accessible in near future. 
It is demonstrated that the efficiency of proton acceleration from relatively thick hydrogen solid ribbon largely exceeds the acceleration efficiency for a thinner ionized plastic foil, which can be explained by enhanced hole boring driven by laser ponderomotive force in the case of light ions and lower target density. 
Three-dimensional (3D) particle-in-cell (PIC) simulations of laser pulse interaction with relatively thick hydrogen target show larger energies of protons accelerated in the target interior during the hole boring phase and reduced energies of protons accelerated from the rear side of the target by quasistatic electric field compared with the results obtained from two-dimensional (2D) PIC calculations. 
Linearly and circularly polarized multiPW laser pulses of duration exceeding 100~fs show similar performance in terms of proton acceleration from both the target interior as well as from the rear side of the target. 
When ultrashort pulse ($\sim$ 30~fs) is assumed, the number of accelerated protons from the target interior is substantially reduced. 
\end{abstract}

\newpage

\section{Introduction}

Ion acceleration driven by high-power femtosecond laser pulses has been attracting great interest for last two decades. 
Most experimental groups have been studying laser-ion acceleration from thin metal or insulator foil targets driven by thermal expansion of laser-heated electrons in the so-called Target Normal Sheath Acceleration (TNSA) mechanism \cite{Macchi2013,Daido2012}. 
However, this mechanism has a limited efficiency, i.e., usually only a few percent of laser pulse energy is transferred into the kinetic energy of accelerated ions which are mostly protons from low-Z hydrocarbon deposits on the target rear surface \cite{Brenner2014}. 
On the path towards increased efficiency of laser-proton acceleration, alternative mechanisms to TNSA have to be investigated. 
One of the most promising alternative mechanisms is radiation pressure acceleration (RPA) \cite{Macchi2013,Daido2012}. 
In this scenario, a compressed cloud of electrons is created by the ponderomotive force driven by incident laser beam in the irradiated layer of the target, electrostatic field arises from this charge separation in plasma and accelerates ions. 

Previous theoretical papers on RPA are mostly devoted to the interaction of ultrashort laser pulses with ultrathin targets (with the thickness below $1~\rm{\mu}m$). 
In this case, mostly RPA in light-sail (LS) regime was studied by various theoretical groups, e.g. \cite{Esirkepov2004,Klimo2008,Robinson2008}. 
However, since there are strict requirements on the laser beam and target quality it will be extremely difficult to demonstrate pure light sail regime in the laboratory even with a higher laser peak power \cite{Macchi2013}. 
On the other hand, RPA in the hole boring (HB) regime \cite{Macchi2005,Liseykina2008,Schlegel2009} should be less sensitive to laser prepulse and spatial intensity variations. 
Moreover, this mechanism should also work efficiently for a thicker targets where the efficiency of other mechanisms such as TNSA is substantially reduced. 
Due to multidimensional effects, generation of hot electrons occurs even for circularly polarized laser beams apart from 1D models \cite{Dollar2012}. 
Thus, one should asses the efficiency of various acceleration mechanisms which take place at the same time. 

The use of newly developed hydrogen solid cryogenic target with the thickness down to a few tens of microns, the so-called thin ribbon of solid hydrogen \cite{Garcia2014}, was demonstrated in experiments with nanosecond laser \cite{Margarone2016} and experiments with these targets on high intensity pico-/femtosecond laser facilities are foreseen. 
The target from ionized solid hydrogen should be relatively thin, low density, capable of producing only protons (with no contaminants) and of operating at a high repetition rate as both refreshable and debris free. 
Thus, this hydrogen solid ribbon is a good candidate for HB RPA regime which requires a relatively low density (but overdense) targets composed of light ions in order to reach a high hole boring (HB) velocity $u_{hb}$ \cite{Qiao2012}. 

In this paper, we demonstrate by multidimensional particle-in-cell simulations using the code EPOCH \cite{Arber2015} that HB RPA mechanism hugely dominates over TNSA both in numbers and maximum energy of accelerated protons for laser power of several PWs and pulse duration of several hundred femtoseconds by using ionized hydrogen ribbon of realistic thickness. 
Such laser beam parameters should be available in near future in the frame of ELI-Beamlines project (L4 laser) \cite{Rus2015}. 
We also investigate the influence of relatively short-scale preplasma and laser wave polarization on the interaction. 

We should note that terminology of the mechanism called HB RPA can be somewhat misleading. 
While LS RPA regime is well established and recognized since the energy of accelerated protons corresponds to the velocity of the accelerated ultrathin foil, the maximum energy of protons in HB RPA regime corresponds to velocity 2$u_{hb}$, thus twice the recession velocity of the plasma surface irradiated by the laser. 
The acceleration of ions can be seen as their reflection from the front of the laser pulse interacting with a homogeneous plasma where plasma-vacuum interface is moving with velocity $u_{hb}$ \cite{Bulanov2012} and is sometimes referred as shock acceleration or piston acceleration \cite{Macchi2013}. 
Throughout this paper, we will use the term HB RPA for all proton acceleration in the target interior during laser-target interaction.

\section{Comparison of hydrogen target with plastic foil}
In order to illustrate the influence of electron density of ionized targets and the target composition on the efficiency of acceleration mechanisms, we assumed $25~\rm{\mu m}$ thick hydrogen ribbon (i.e., foil in the following discussion) and $5~\rm{\mu m}$ thick fully ionized polyethylene (CH2) foil of realistic target density irradiated by the most intense laser pulse assumed in the frame of this paper (with peak intensity $3 \times 10^{22}~\rm{W/cm^2}$, dimensionless amplitude $a_0 \approx 163$). 
Taking into account electron densities $n_e$ for both targets ($56~n_{ec}$ vs. $339~n_{ec}$, where $n_{ec}$ is the critical density) it implies that areal electron densities of the targets $n_e l$ (where $l$ is the target thickness) are roughly the same.

Two different setups are used for 2D simulations with fully ionized hydrogen and polyethylene foils. 
In the case of simulations with hydrogen target, the laser plasma interaction occurs in the simulation box with dimensions $150~\rm{\mu m} \times 50~\rm{\mu m}$, which contains square cells with the size of 20~nm. 
Due to significantly higher electron density and therefore higher requirements on simulation stability, the cell size is shorten to 10~nm in the case of polyethylene target and simulation box is reduced to $110~\rm{\mu m} \times 50~\rm{\mu m}$. 
Each cell occupied by plasma contains 56 electrons in the case of hydrogen target and 339 electrons in the case of polyethylene target. 
Both targets are placed $50~\rm{\mu m}$ from the simulation box boundary with the entering laser pulse at position $x = 0$, i.e., the laser pulse front reaches the edge of the plasma foil at the same time, referred as $t = 0$, in all simulations. 
The transverse size of the targets is $50~\rm{\mu m}$, i.e., the targets are touching simulation box boundaries at positions $y = \pm 25~\rm{\mu m}$ where thermal boundary conditions for particles are applied. 
The used laser pulses incident normally on the target and have a $\sin^2$-function time profile whereas the spatial profile is Gaussian. 
The full pulse duration is 320~fs (then, the pulse energy is equal and the pulse temporal profile is similar to gaussian profile with FWHM equal to 150~fs). 
Laser beam width is $5~\rm{\mu m}$ at FWHM, which implies laser peak power about 9~PW for the assumed intensity $3 \times 10^{22}~\rm{W/cm^2}$. 

Despite its relatively low density (less than $a_0 n_{ec}$), the cryogenic hydrogen target does not become transparent to the simulated laser pulse. 
In reality, the relativistic transparency threshold is much higher than at $a_0 n_{ec}$ given by linear analysis \cite{Mora2001} as the laser pulse strongly compress the electron fluid, raising the electron density at the pulse front considerably \cite{Cattani2000}. 
Thus, HB RPA can occur even well above the intensity assumed here \cite{Robinson2012}. 
With the use of high intensities (above $10^{22}~\rm{W/cm^2}$) the RR force can appreciably affect proton acceleration \cite{Tamburini2012}, therefore we studied both cases with QED module switched on in EPOCH code \cite{Arber2015}. 

\begin{figure}[h]
\begin{center}
\includegraphics[width=0.45\textwidth]{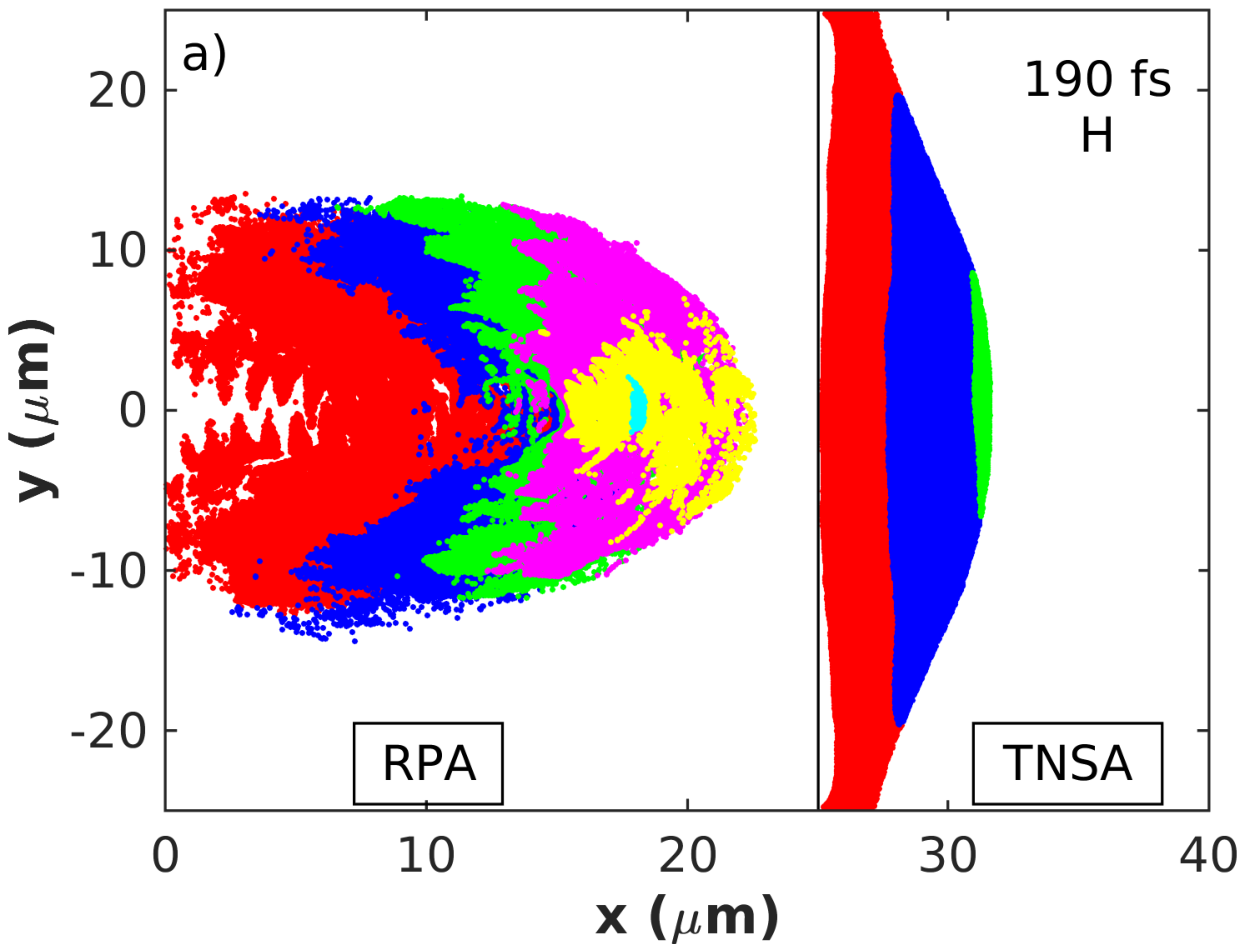}
\includegraphics[width=0.45\textwidth]{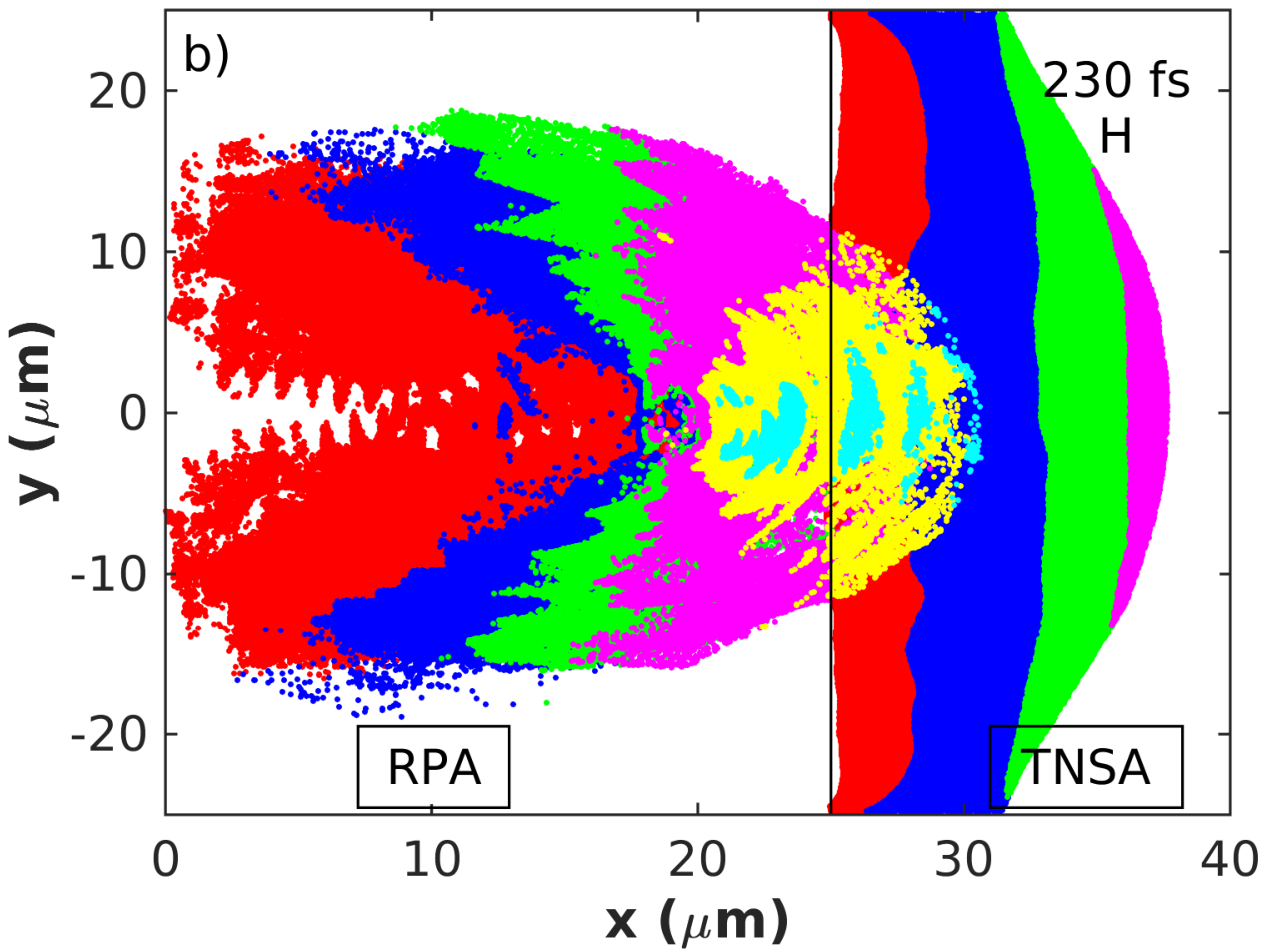}
\includegraphics[width=0.45\textwidth]{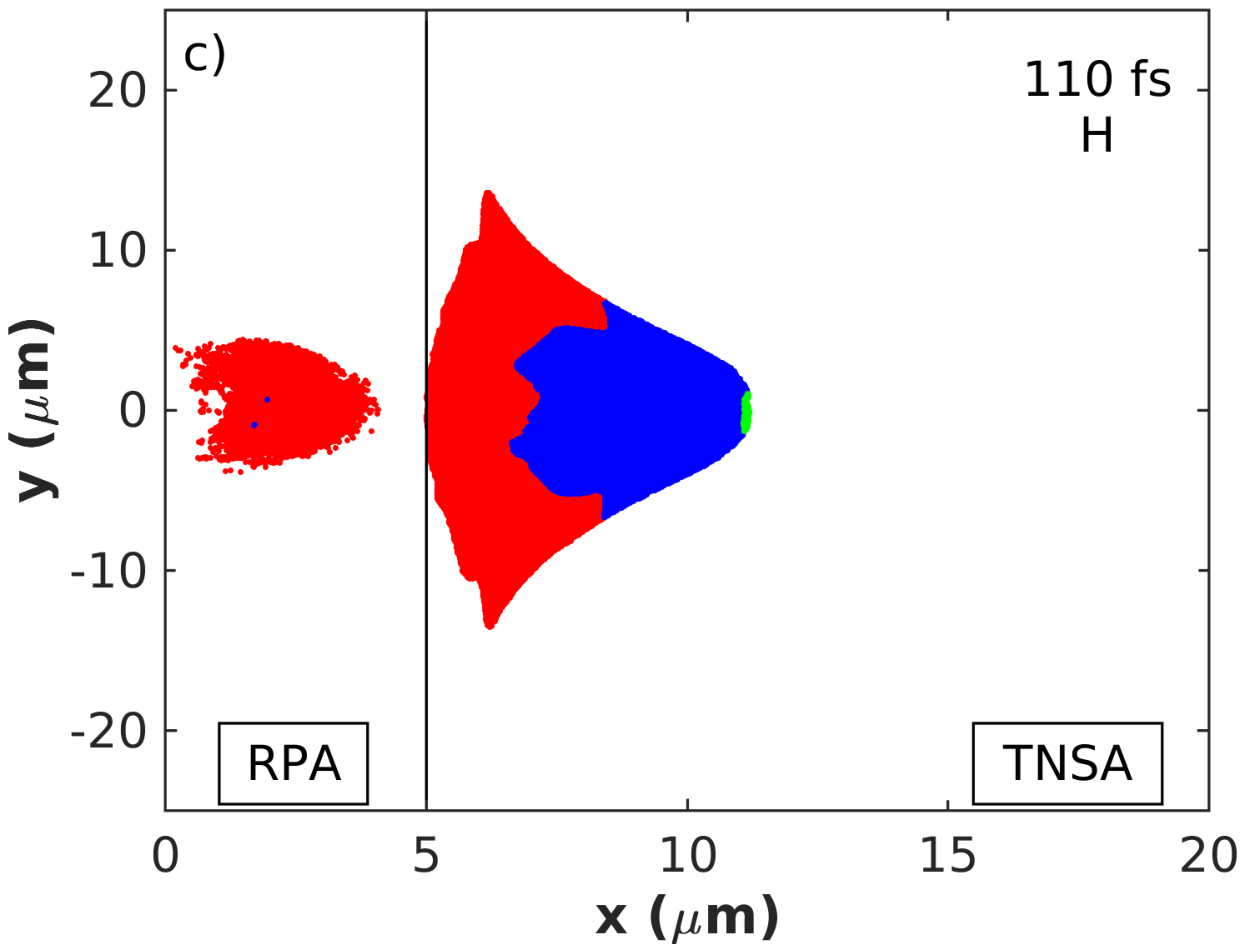}
\includegraphics[width=0.45\textwidth]{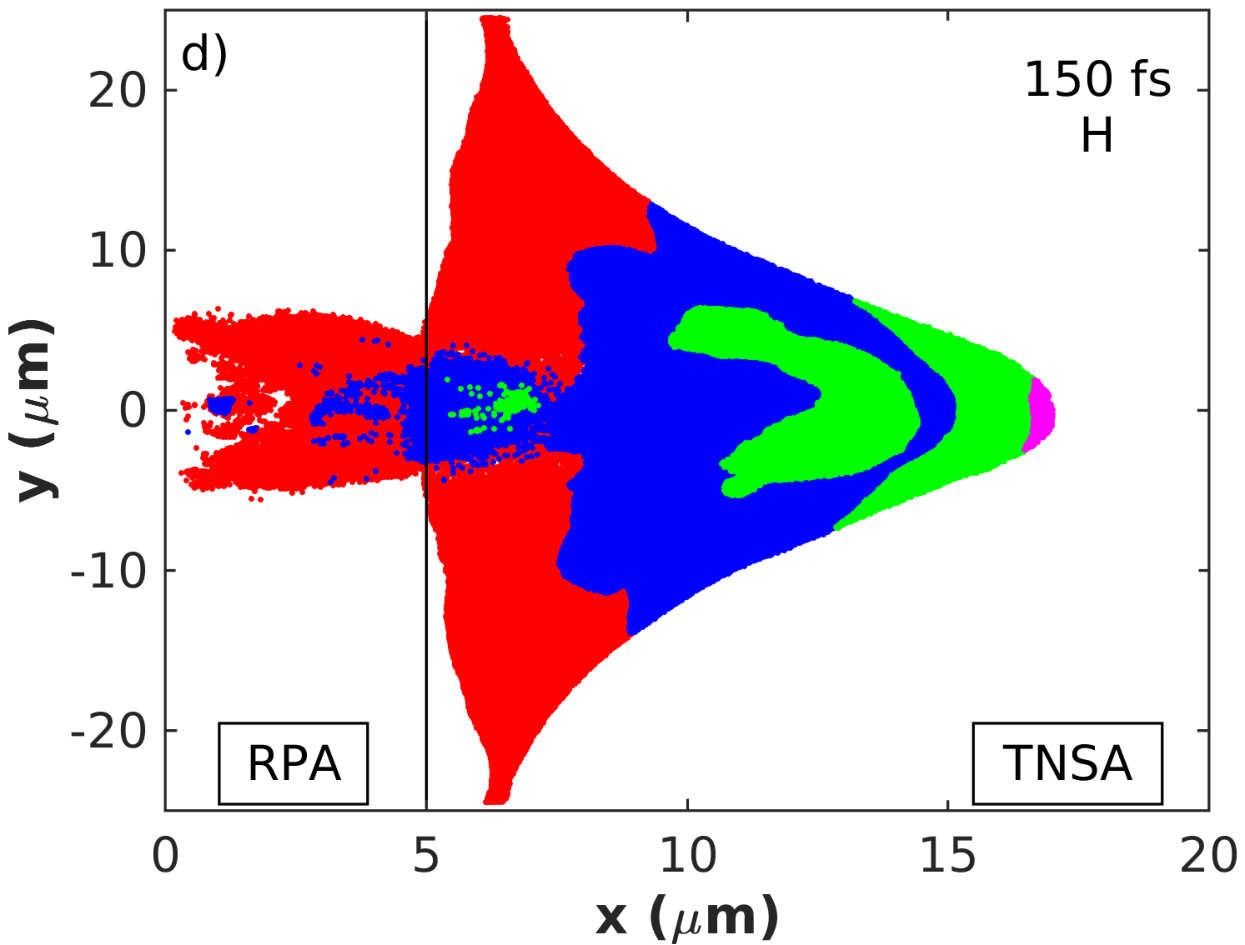}
\includegraphics[width=0.9\textwidth]{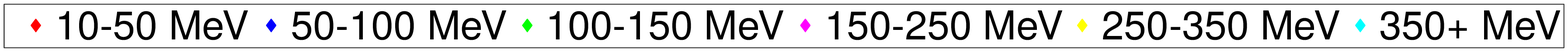}

\caption{\label{fig1} (a), (b) Proton energy layers for hydrogen ribbon at 190~fs and at 230~fs from the beginning of laser-target interaction; (c), (d) Proton energy layers for polyethylene target at 110~fs and at 150~fs, respectively, from the beginning of laser-target interaction.
The targets are initially located between $x=0$ and $25~\rm{\mu m}$ or between $x=0$ and $5~\rm{\mu m}$.}
\end{center}
\end{figure}

Since CH2 target is composed of heavier ions than only protons and its ion density is larger compared with hydrogen target, we can expect higher hole boring velocity $u_{hb}$ for the hydrogen. 
Indeed, our analysis of simulation data gives $u_{hb} = 0.08~c$ for CH2 target and $0.31~c$ for hydrogen ribbon. 
Thus, we can expect much higher energies of protons accelerated by HB RPA in the case of hydrogen taking into account that the maximum energy of such protons is roughly proportional to $u_{hb}$ squared (neglecting relativistic effects). 

\begin{figure}[h]
\begin{center}
\includegraphics[width=0.45\textwidth]{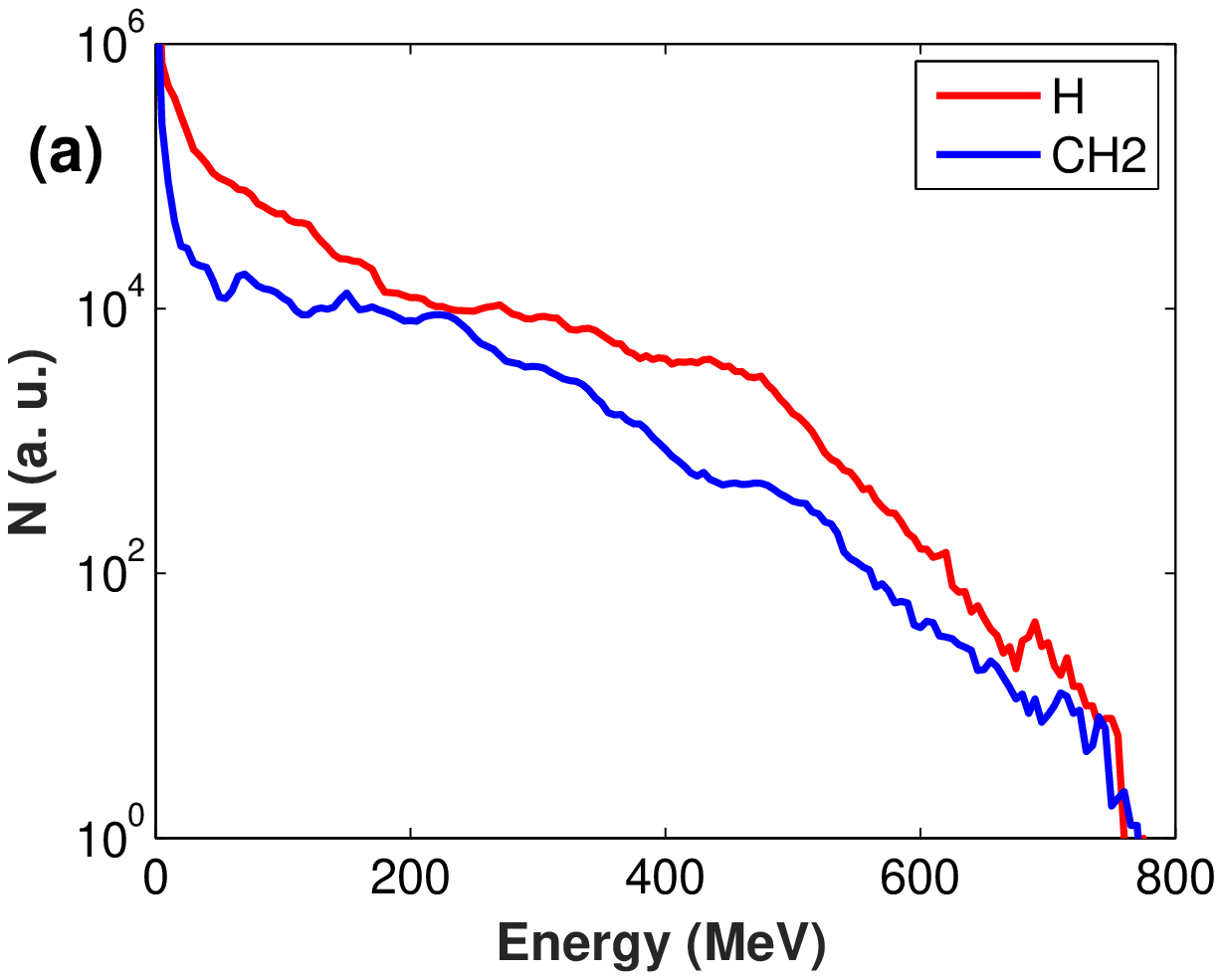}
\includegraphics[width=0.45\textwidth]{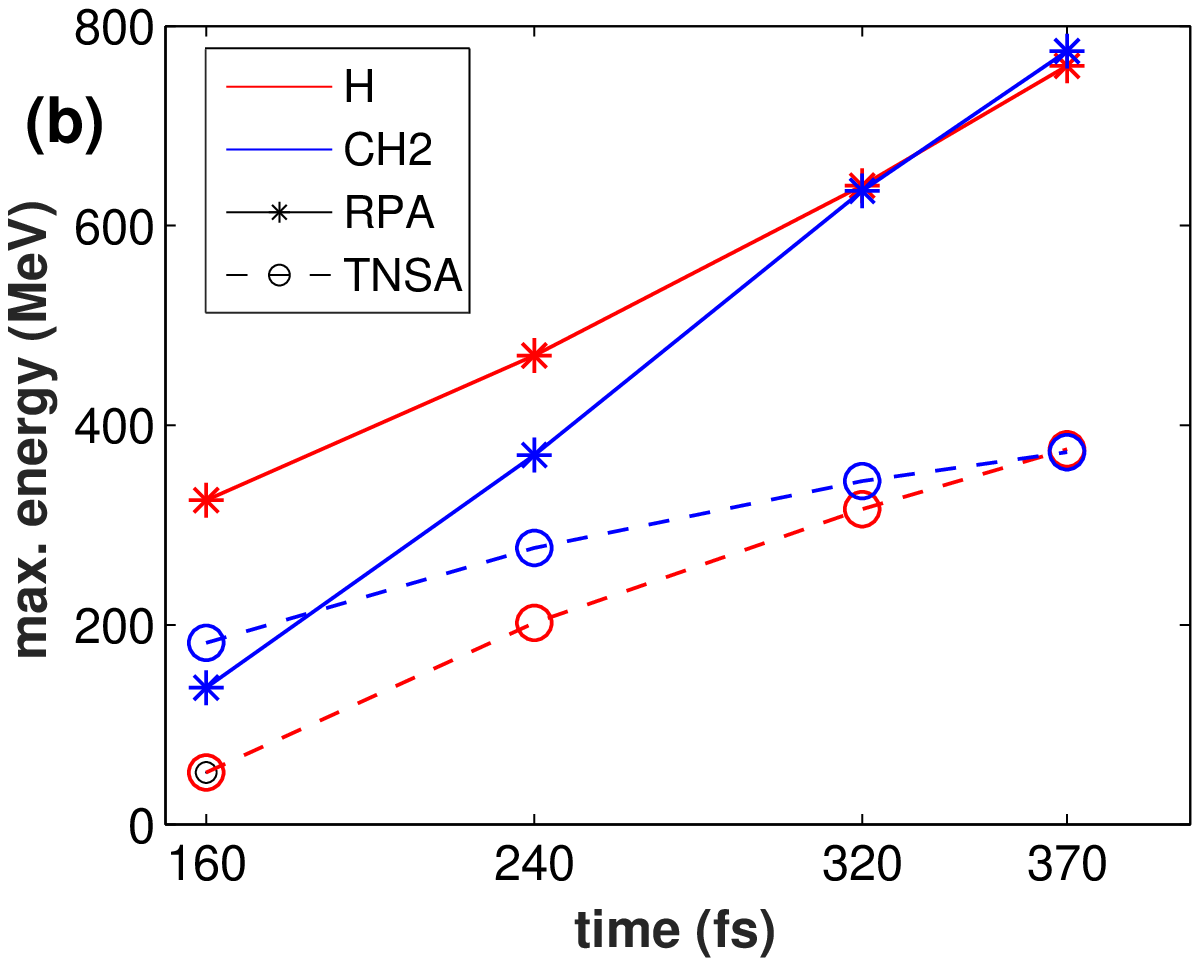}

\caption{\label{fig2} (a) Proton energy spectra at 370~fs from the beginning of laser-target interaction; 
(b) Temporal evolution of maximum proton energies for hydrogen (H) and polyethylene (CH2) targets. 
The protons marked as "RPA" are initially accelerated by HB RPA mechanism and originate from the target interior (however, they cross the initial rear side target position later in time as shown in Fig. \ref{fig1}). 
The protons marked as "TNSA" originate from the rear side of the target. }
\end{center}
\end{figure}

Fig. \ref{fig1} shows the presence of protons at various energy intervals in the simulation area (proton energy layers). 
The acceleration of protons in targets can be divided into three stages. 
In the first stage shown in Fig. \ref{fig1} a), c), there are two groups of protons which can be clearly distinguished from each other in space: the protons accelerated inside the target in the laser pulse propagation direction by HB RPA and the protons accelerated by TNSA from the rear side of the target. 
In the second stage shown in Fig. \ref{fig1} b), d), the most energetic protons accelerated to velocities higher than $u_{hb}$ inside the target enters into the TNSA field behind the initial position of the target rear side, however, the target is still not transparent for the laser pulse. 
If the laser pulse punched through the target, the third stage occurs when the protons can be further accelerated to very high energies. 
In this transparency phase, the laser continuously imparts forward momentum to the electrons, which couple to the ions \cite{Jung2013a,Liu2013}. 
This third stage, which is sometimes called break-out afterburner (BOA) mechanism \cite{Yin2011,Jung2013b}, is more efficient in the case of CH2 target due to an earlier punching of the laser through the target (i.e., at 200 fs for CH2 target vs. at 270 fs for H ribbon), closer to the laser pulse maximum amplitude as well as larger plasma density allowing efficient volumetric absorption of laser pulse energy to electrons for longer time. 
That is why we can observe similar maximum energies of accelerated protons from both targets at the end of simulation (Fig. \ref{fig2} a)), even though the observed energies are initially larger for hydrogen target as can be seen in Fig. \ref{fig2} b). 
Nevertheless, the number of accelerated protons is substantially higher for the hydrogen target (about 4~times integrated all protons with energy exceeding 10~MeV) as shown in the energy spectra in Fig. \ref{fig2} a). 
The conversion efficiency of laser pulse energy to high-energy protons (for those exceeding 10~MeV) is about 9~percent for CH2 target and 27~percent for hydrogen ribbon. 
Thus, we have shown that even several times thicker hydrogen foil can be advantageous for proton acceleration compared with a standard target usually used in the experiment.

\section{2D simulations at various laser intensities}
In order to assess the efficiency of HB RPA mechanism for novel hydrogen target, it would be worth to study the interaction at lower peak intensities than maximum $3 \times 10^{22}~\rm{W/cm^2}$ which will be very challenging to reach under real experimental conditions. 
Therefore, we reduced the intensity (and energy) of the pulse by factors 1/2, 1/4 and 1/8 in the following calculations (thus, we assumed laser power about 4.5, 2.25 and 1.1~PW on target, respectively). 

\begin{figure}[h]
\begin{center}
\includegraphics[width=0.55\textwidth]{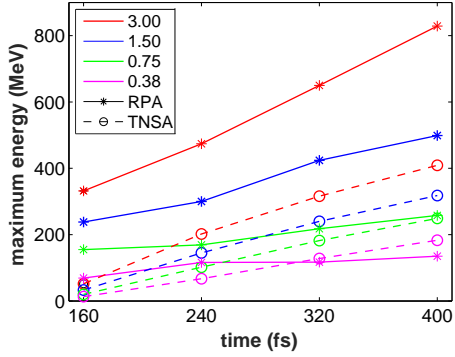}

\caption{\label{fig3} Temporal evolution of maximum energies of protons initially accelerated in the interior of hydrogen ribbon by HB RPA mechanism and of protons accelerated by pure TNSA mechanism for various laser intensities ranging from $3.75 \times 10^{21}$ to $3 \times 10^{22}~\rm{W/cm^2}$.}
\end{center}
\end{figure}

In order to summarize the results for the assumed laser intensities, we start with the temporal evolution of maximum energies of protons accelerated by two mechanisms shown in Fig. \ref{fig3}. 
Our simulations reveal that the pulse burns through the target for laser intensity equal or larger than $1.5 \times 10^{22}~\rm{W/cm^2}$. 
Otherwise, the laser and target parameters are not sufficient to reach this stage of the interaction. 
Such behavior affects later stage of the acceleration of protons originated from the target interior when their maximum energies are rapidly enhanced for higher intensities (see solid lines in Fig. \ref{fig3}), whereas in the case of laser intensity below $1.5 \times 10^{22}~\rm{W/cm^2}$ mixture RPA/TNSA mechanism accelerates protons originated from the target interior to energies similar to the protons originated from the target surface (accelerated by pure TNSA). 
In the case of the lowest intensity assumed here ($3.7 \times 10^{21}~\rm{W/cm^2}$), mixture RPA/TNSA mechanism almost did not occur (protons originated from the target interior and initially accelerated by HB RPA are only slightly post-accelerated in the TNSA field) and TNSA mechanism accelerates protons to the highest energies. 

Additional analysis of this set of simulations have shown us that the number of protons initially accelerated from the target interior ("RPA" protons) largely exceeds the number of protons accelerated by pure TNSA mechanism ("TNSA" protons). 
The ratio of these two kinds of accelerated protons together with the energy transformation efficiency of the laser pulse to all protons with kinetic energy exceeding 10~MeV accelerated in the forward direction is shown in Table \ref{tab1}. 
Generally, the number of protons initially accelerated by HB~RPA is increasing with laser intensity. 
Even though we cannot well determine the numbers at later stage after the time instant when RPA protons reach the TNSA field, the ratio of RPA protons to TNSA protons is higher in all studied cases before these two groups are mixed together: it is about 2 for the lowest intensity case and reaches about 9 for the highest intensity case at the end of the first stage of proton acceleration. 
Energy transformation efficiency of the laser pulse energy to high-energy protons ($> 10~\rm{MeV}$) is also enhanced from 16~percent at $3.7 \times 10^{21}~\rm{W/cm^2}$ to 27 percent at $3 \times 10^{22}~\rm{W/cm^2}$. 
While the former efficiency is at the borders what was achieved experimentally \cite{Brenner2014}, the latter is far beyond the current observations due to RPA dominant mechanism and pure hydrogen target. 
Here, the efficiency of TNSA mechanism is reduced by the spread of hot electrons towards lateral sides of the target which is not the case for HB RPA driven by the ponderomotive force. 

\begin{table}[!h]
\caption{\label{tab1} Energy transformation efficiency of laser pulse energy to kinetic energy of fast protons (only protons with energy $\varepsilon_{k} > 10~\rm{MeV}$ are taken into account) and the ratio of the number of fast protons accelerated in the target interior (RPA protons) to the number of protons accelerated from the target rear side (TNSA protons) for laser intensities ranging from $3.7 \times 10^{21}$ up to $3.0 \times 10^{22}~\rm{W/cm^2}$. 
The ratio of fast protons is determined at time moment specified in round brackets since the protons accelerated by HB RPA in the target interior propagate through initial position of the target's rear side and cannot be well distinguished from TNSA protons afterwards. 
}
\begin{center}
\begin{tabular}{|l|l|l|} \hline

laser intensity                 &  energy transformation & \#RPA / \#TNSA protons \\
($\times 10^{22}~\rm{W/cm^2}$)  &  efficiency            & ($\varepsilon_{k} > 10~\rm{MeV}$)        \\
\hline
\hline
3.0 & 0.27 & 9.0 (190~fs)  \\ \hline
1.5 & 0.27 & 6.3 (220~fs)  \\ \hline
0.75 & 0.21 & 3.6 (260~fs)  \\ \hline
0.375 & 0.16 & 2.0 (320~fs) \\ \hline
\end{tabular}
\end{center}
\end{table}

We should note that the protons accelerated by HB RPA mechanism should travel through the target interior. 
Since hydrogen ribbon can be relatively thick target, the accelerated protons can collide with other particles in the target and be slowed down and scattered. 
However, these effects should be negligible taking into account that mass stopping power of 10~MeV protons in hydrogen is $103~\rm{MeV/(g/cm^2)}$ and the stopping power is decreasing with increasing proton kinetic energy. 
Thus, the average loss of kinetic energy is less than 0.1 MeV per $100~\rm{\mu m}$ for protons accelerated to energy $>10~\rm{MeV}$.

\section{Influence of initial density profile}
The requirements on the intensity contrast of laser pulse of multiPW power in order to protect the target from heating, melting and evaporation before the arrival of the main pulse are even more challenging than for current laser facilities. 
Even if the contrast will become comparable to the best contrast values presently achieved at subPW lasers, one can expect that the target will be partially heated before the main pulse and preplasma formation will occur. 
Therefore, it is important to investigate how the preplasma will affect the whole interaction and proton acceleration. 

\begin{figure}[h]
\begin{center}
\includegraphics[width=0.45\textwidth]{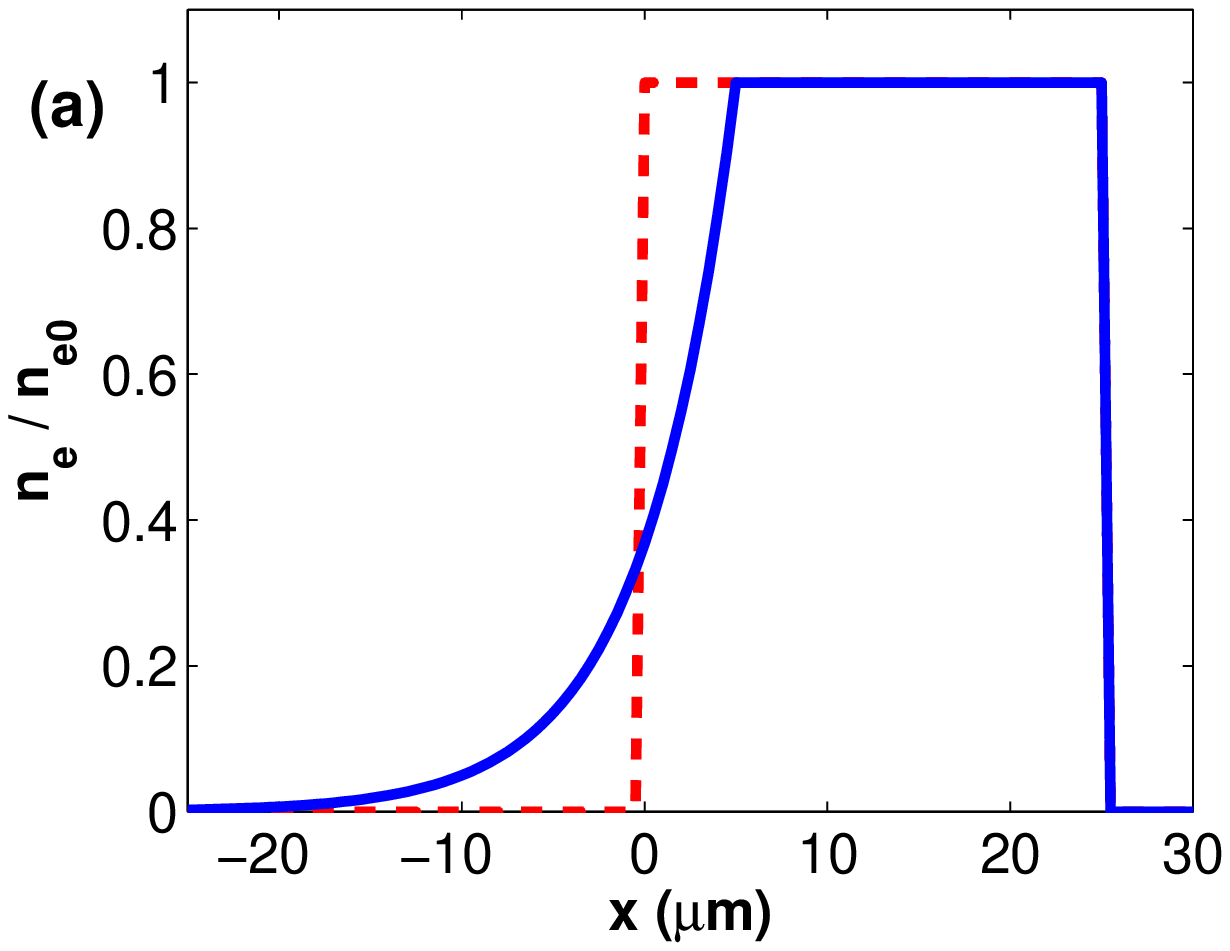}
\includegraphics[width=0.45\textwidth]{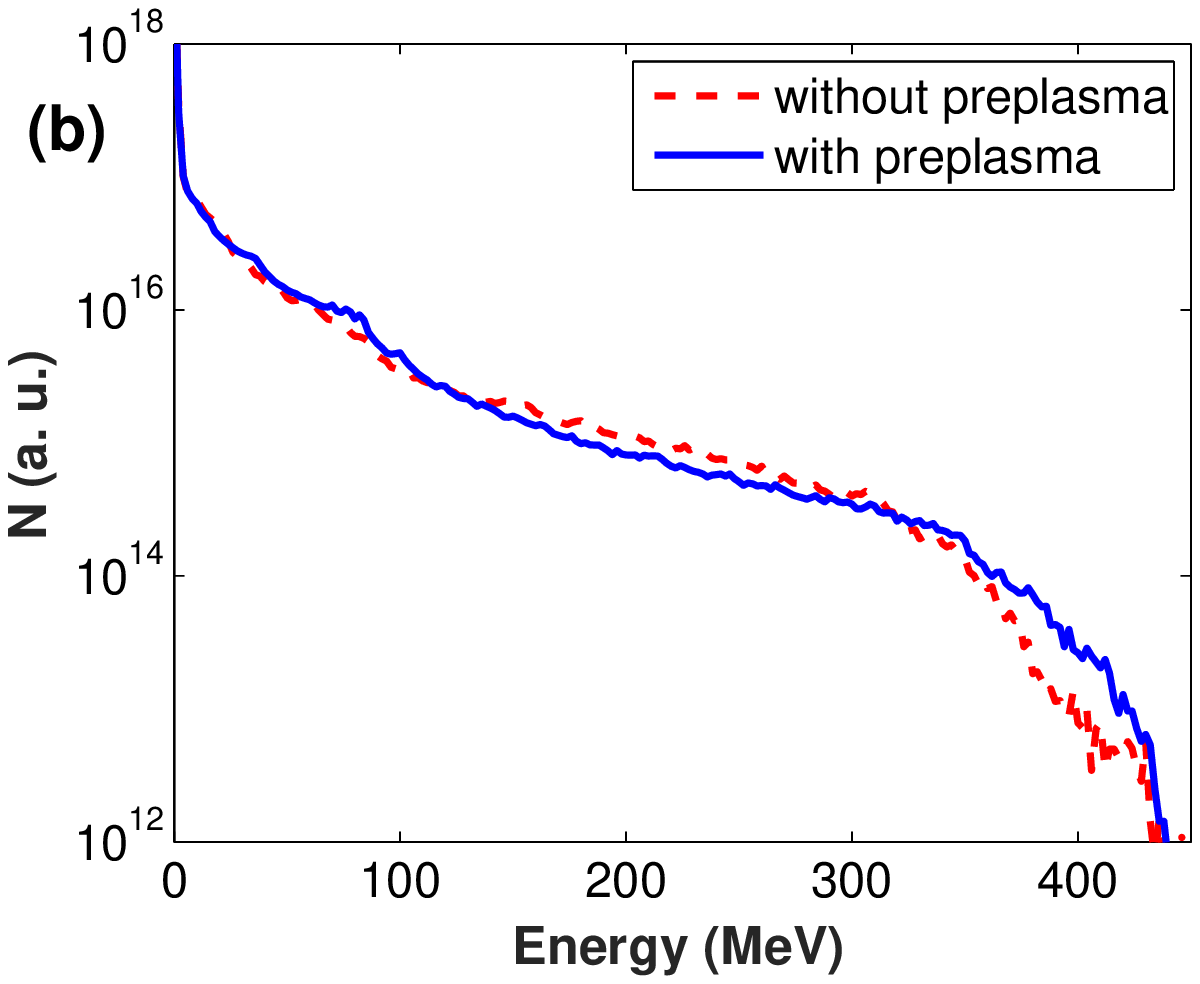}

\caption{\label{fig4} (a) Initial density profile along longitudinal spatial axis (laser propagation direction) normalized to maximum density in the case of target with preplasma (solid line) and without preplasma (dashed line); (b) Final energy spectra of accelerated protons for simulations of hydrogen target irradiated by the laser pulse with/-out preplasma. }
\end{center}
\end{figure}

In our additional 2D simulations, we assumed exponential density profile of preplasma $\exp(-x/L_{sc})$ for simplicity as in many papers (e.g. \cite{Orban2015,Loch2016}) with the scale length $L_{sc}=5~\rm{\mu m}$. 
We did not assume multidimensional effects in the preplasma formation \cite{Esirkepov2014}, so the density of preplasma depends only on the longitudinal ($x$) axis in the simulation box. 
Since the number of macroparticles is the same in all cells occupied by plasma, we have to take into account numerical weightings of macroparticles in the following analysis. 
1D profile of target density was set in order to keep constant the number of real particles in the target, which means that the thickness of maximum density region was decreased from $25~\rm{\mu m}$ to $20~\rm{\mu m}$ and the length of preplasma region was set to $30~\rm{\mu m}$, see Fig. \ref{fig4} a). 
Laser peak intensity is assumed at $1.5 \times 10^{22}~\rm{W/cm^2}$. 

For our parameters, we observed that the assumed preplasma only slightly modify the final proton energy spectrum (shown in Fig. \ref{fig4} b)) and also the transformation efficiency of laser pulse energy to high-energy protons is similar for the case with and without the preplasma (29~percent vs. 27~percent, respectively). 
We also observed the same ratio of RPA to TNSA protons for both cases. 

\section{3D simulations and comparison with 2D}
Previous studies have shown that 2D PIC simulations can overestimate the energies of accelerated protons \cite{dHumieres2013,Liu2013} or they can capture different physics depending on laser wave polarization \cite{Stark2017}.
Therefore, we performed very demanding 3D calculations in order to check the validity of obtained results in 2D PIC simulations and to analyze the influence of laser wave polarization. 
Due to computational constraints, we assumed reduced thickness of the target and reduced length of the laser pulse which were both decreased to about 3/5 of the initial values down to $15~\rm{\mu m}$ thick hydrogen ribbon and 200~fs full duration of the laser pulse. 
Such parameters are still sufficient to observe the evolution of laser-target interaction and to assess the efficiency of HB RPA and TNSA regimes. 

The interaction of laser pulse with the target has been studied in the simulation box with dimensions $100~\rm{\mu m} \times 34.5~\rm{\mu m} \times 34.5~\rm{\mu m}$. 
The simulation box consisted of cells with the size of 20~nm in the longitudinal direction (in the direction of laser beam propagation) and 30~nm in both transverse directions. 
10~electrons and 10~protons were initialized in each cell occupied by plasma. 
Thus, around $6.6 \times 10^{9}$  cells and $20 \times 10^{9}$ particles were involved in the simulation. 
The simulations were run on 1920 CPU cores for about 100 hours each on Salomon cluster in the Czech Republic (IT4Innovations project). 
We studied the interaction with linearly and circularly polarized laser pulses of peak intensity $1.5 \times 10^{22}~\rm{W/cm^2}$. 
Since linearly polarized laser pulses are mostly used in experiments (that is why we assumed linearly p-polarized pulses in previously described 2D calculations), we start our discussion by comparison of the results from 3D and 2D calculations with linear polarization and with other similar parameters. 
Thus, we performed additional 2D~simulations with reduced target thickness and laser pulse duration as described above. 

\begin{figure}[h]
\begin{center}
\includegraphics[width=0.45\textwidth]{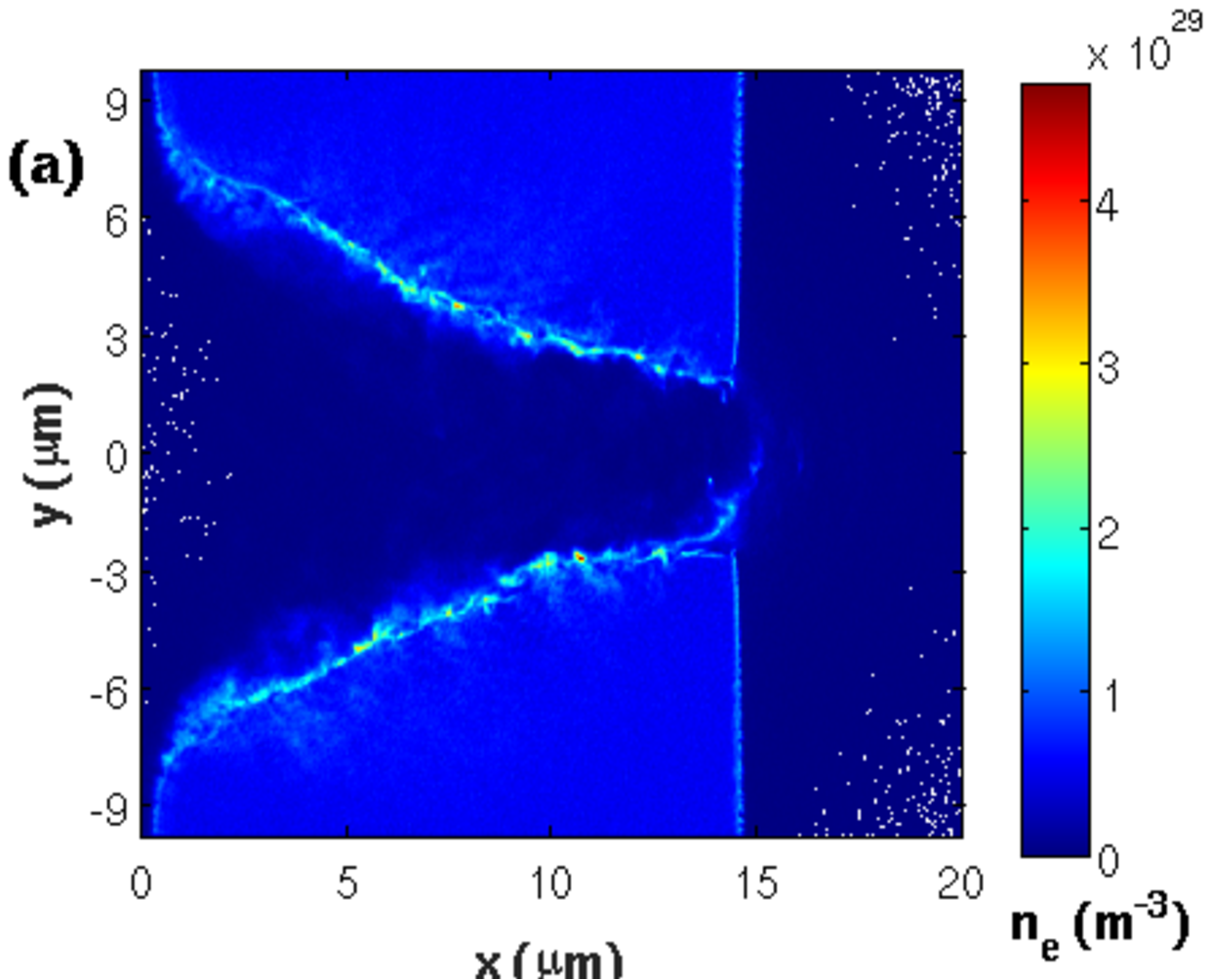}
\includegraphics[width=0.45\textwidth]{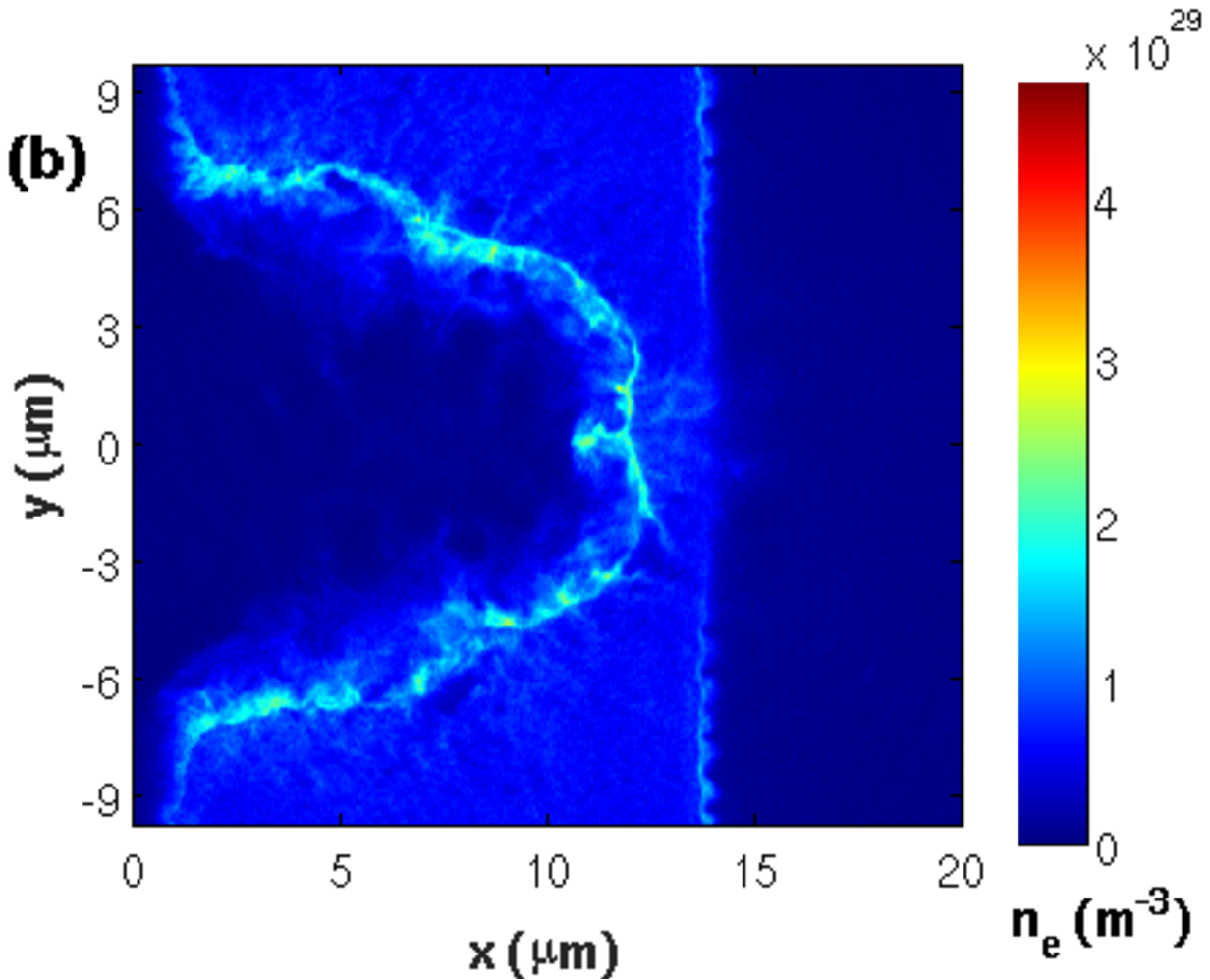}
\includegraphics[width=0.45\textwidth]{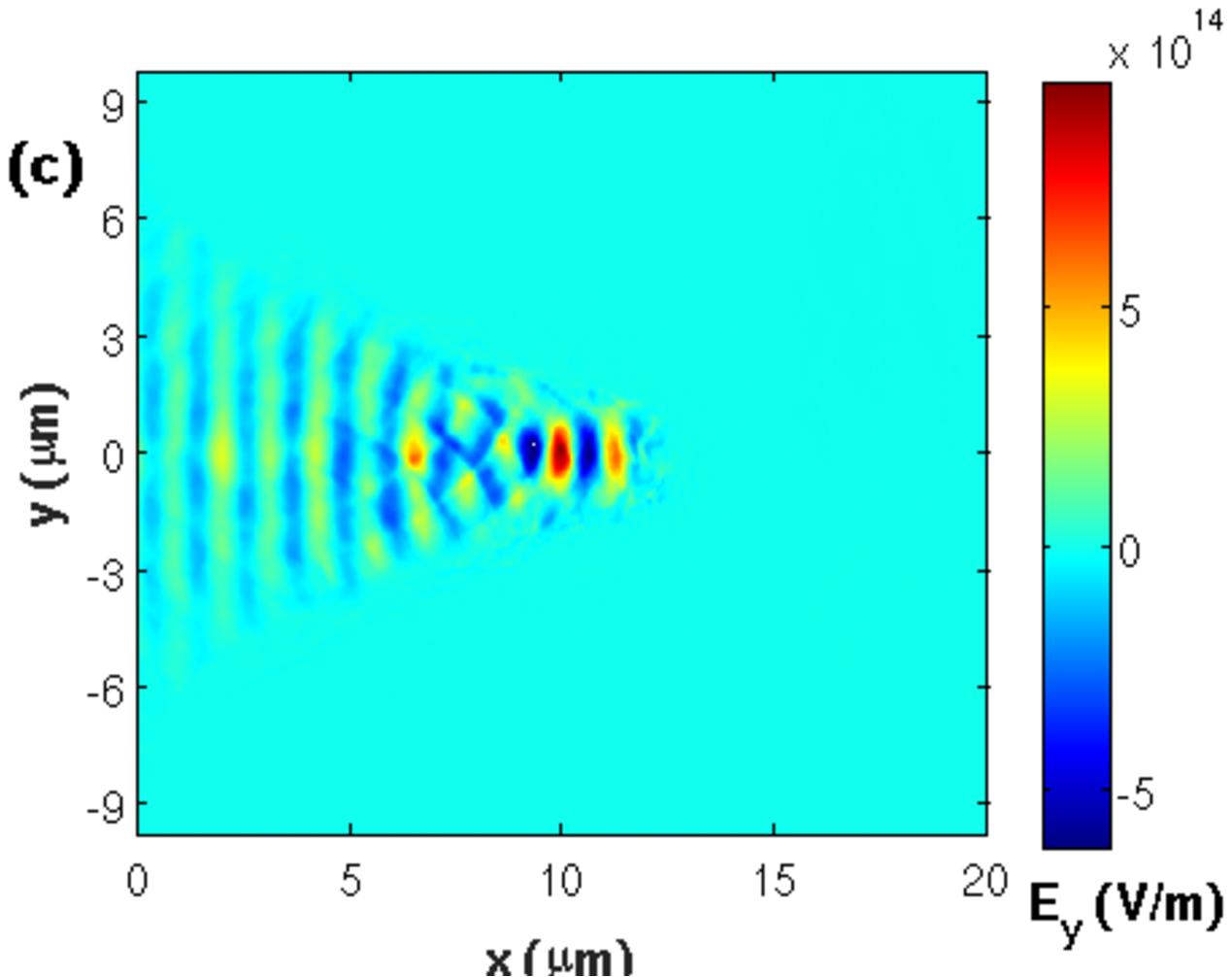}
\includegraphics[width=0.45\textwidth]{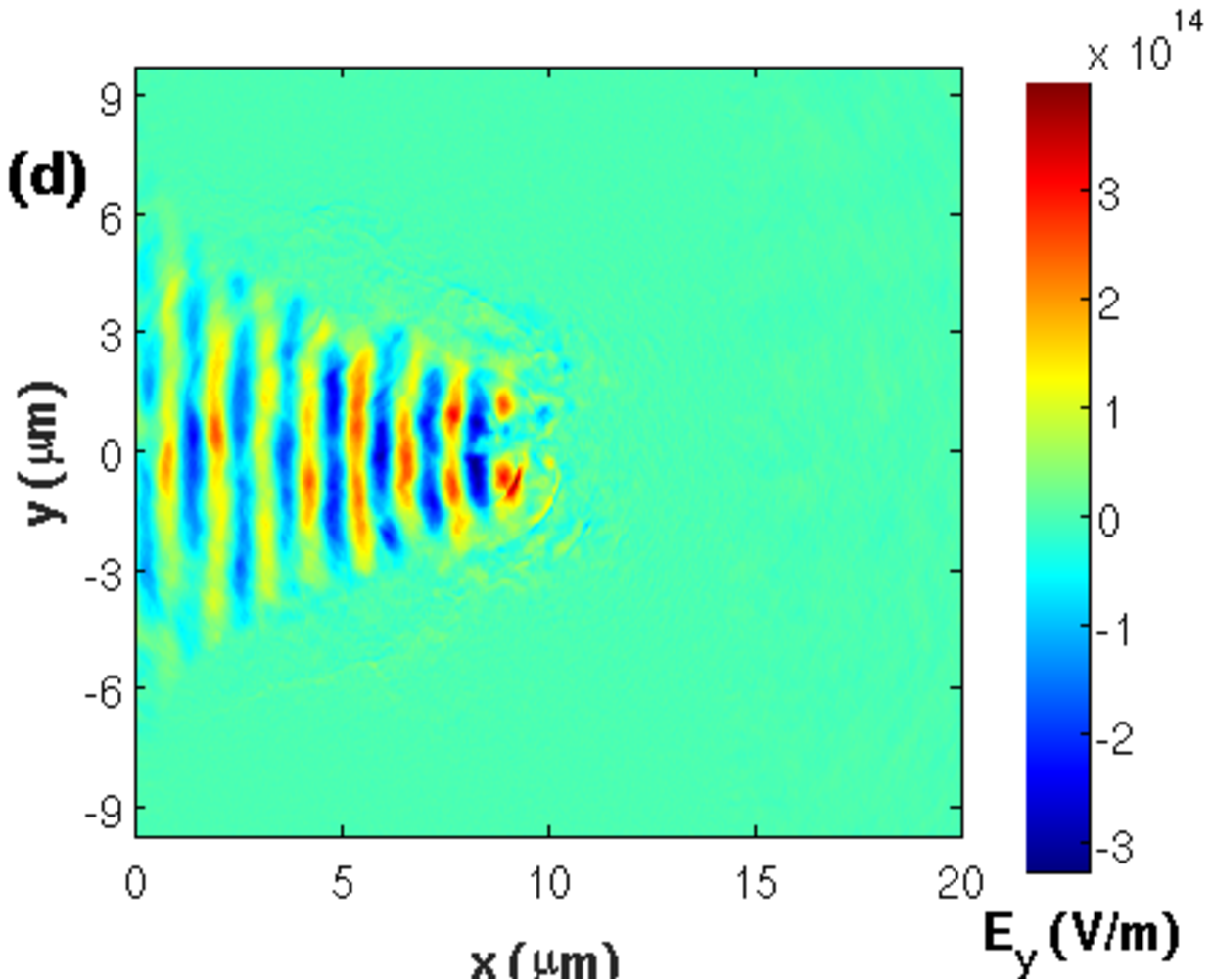}

\caption{\label{fig5} (a), (b) Electron density distribution in 3D and 2D simulations, respectively, at 220~fs from the beginning of laser-target interaction; 
(c), (d) Transverse electric field distribution in 3D and 2D simulations, respectively, at 180~fs from the beginning of laser-target interaction. 
In 3D case, the distributions are taken in $(x, y)$ plane at $z=0$. 
}
\end{center}
\end{figure}

\begin{figure}[h]
\begin{center}
\includegraphics[width=0.45\textwidth]{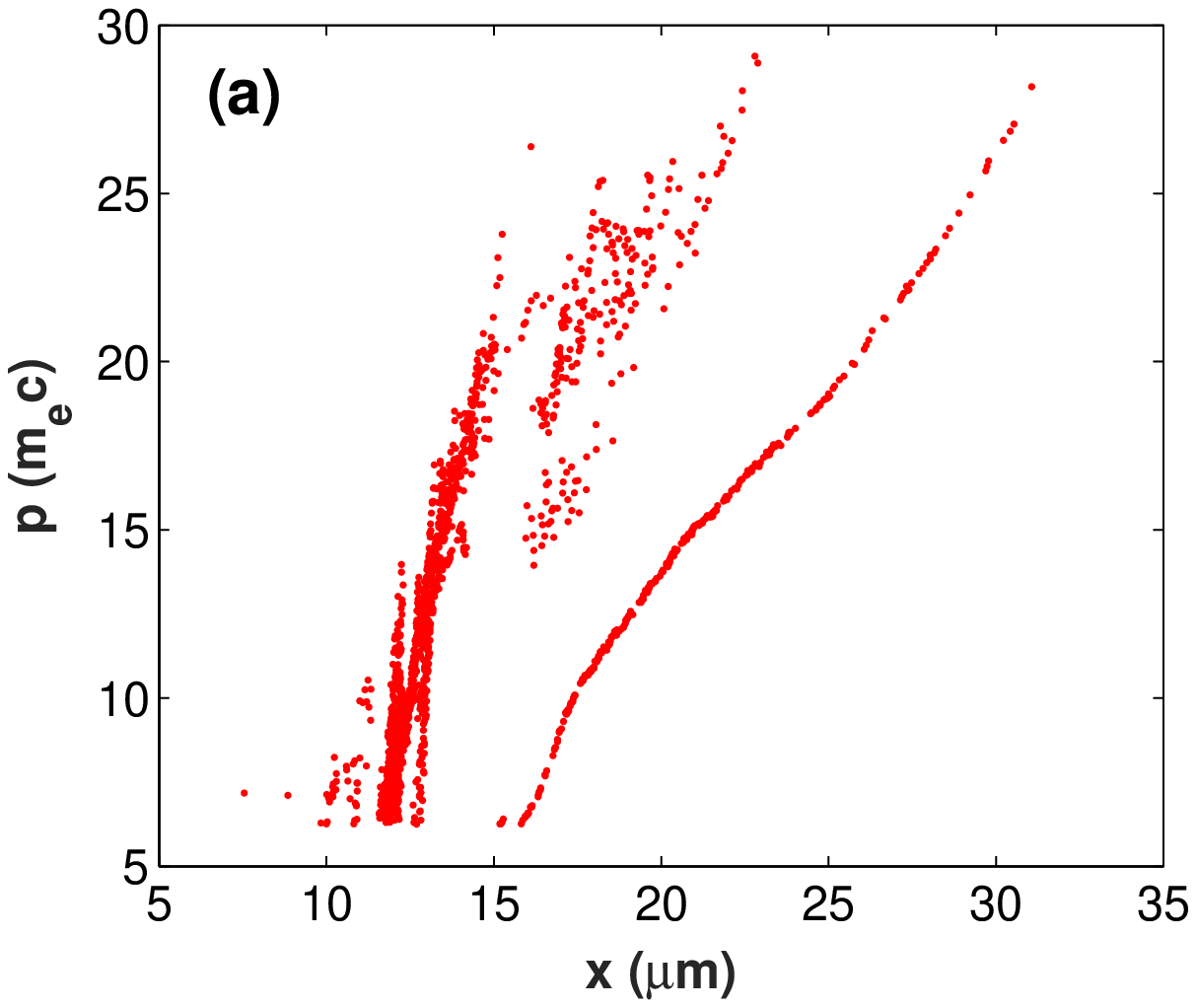}
\includegraphics[width=0.45\textwidth]{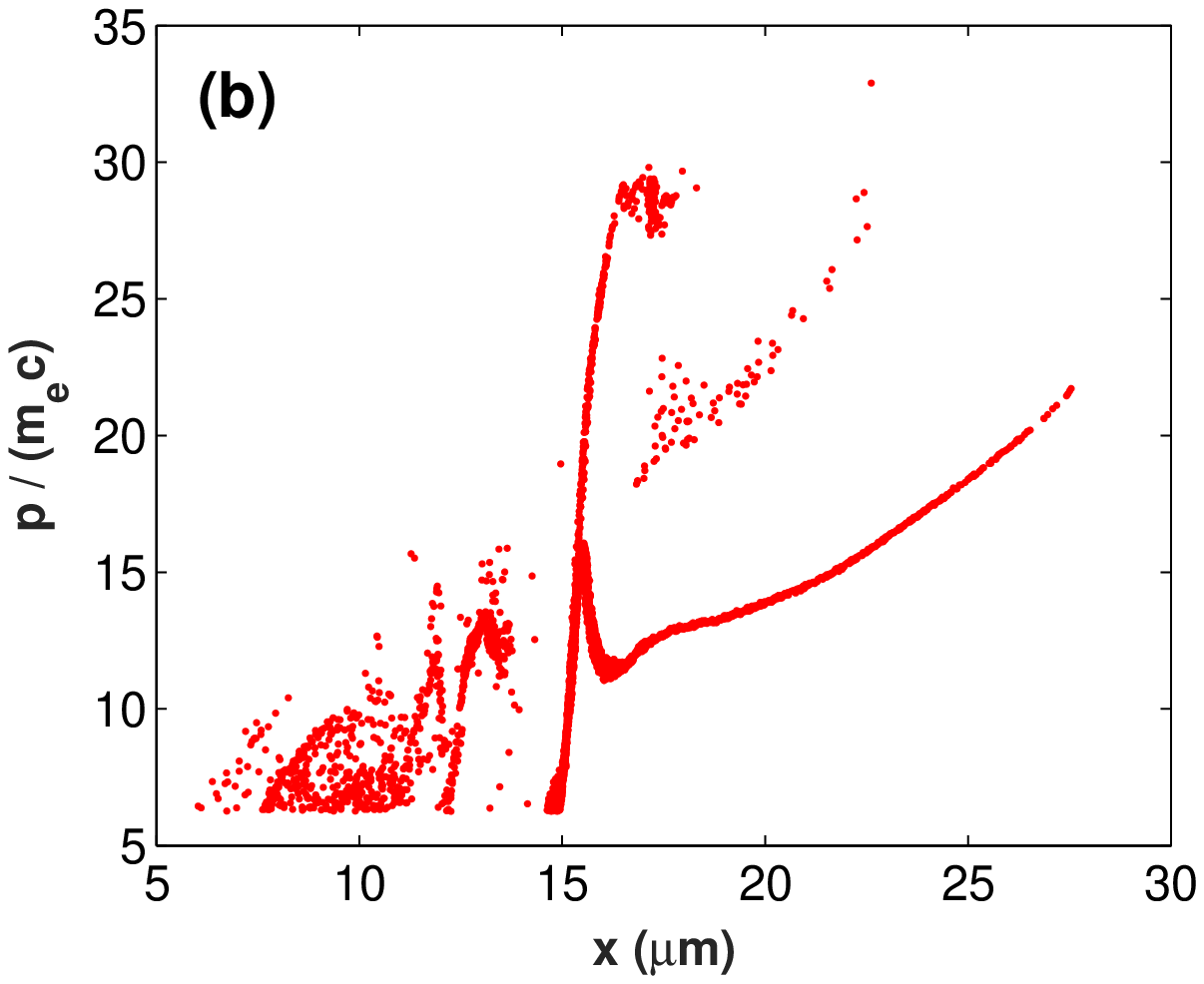}
\includegraphics[width=0.45\textwidth]{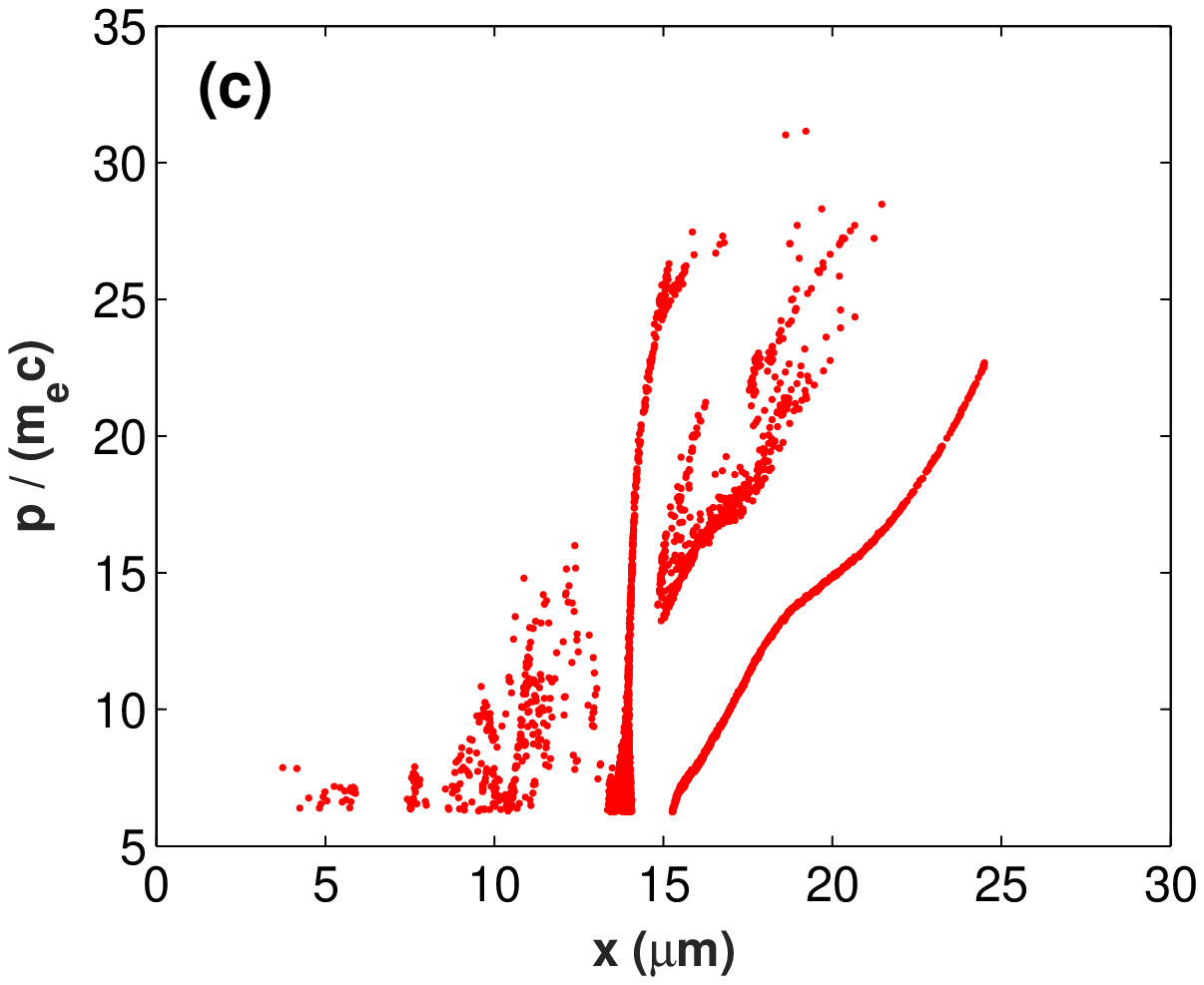}

\caption{\label{fig6} Phase space of protons $(p, x)$, where $p$ is the momentum of protons accelerated in the forward direction with energy $> 10~\rm{MeV}$ for (a) 2D and (b) 3D simulations in the case of linearly p-polarized laser beam, for (c) 3D simulation in the case of circularly polarized beam.}
\end{center}
\end{figure}

When these results from 3D and 2D were compared, we observed several differences: 1) the laser pulse burned through the target during the interaction in the case of 3D whereas it was not the case for 2D; 2) the temperature of hot electrons was higher in 2D simulations (44 MeV in 2D vs. 34 MeV in 3D); 3) the energy of accelerated ions by HB RPA mechanism was higher in 3D whereas the energy of ions accelerated by TNSA was larger in 2D case. 
For example, the maximum energy of protons accelerated by TNSA was about 155~MeV in 2D vs. 80~MeV in 3D at 180~fs, the energy of protons accelerated by HB RPA was about 180~MeV in 2D vs. 250~MeV in 3D at the same time instant. 
Such difference is also apparent in phase spaces of protons in Fig. \ref{fig6} a), b) at the end of laser-plasma interaction. 
Here, the maximum value of momentum reaches almost $30~m_ec$ for both types of protons in 2D~case, whereas it exceeds $30~m_ec$ for RPA protons and slightly above $20~m_ec$ for TNSA protons in 3D~case. 

The observation 1) implies a higher hole boring velocity in 3D~PIC simulations. 
Such observation is in agreement with higher intensity of the focused laser beam in 3D~geometry when the beam burns the hole inside the ionized target. 
In 2D~geometry, the intensity of the focused beam is less enhanced as the beam is focused only in one transverse direction. 
The effect of beam focusing can be observed in Fig. \ref{fig5} showing instantaneous transverse electric fields and electron densities at later stage of laser-plasma interaction. 
From those fields and densities, one can also infer faster narrowing of the beam in 3D~geometry when it penetrates through the target as the intensity gradient in the perpendicular direction to the beam propagation is higher. 

The observation 2) is connected with geometrical effect which leads to the heating of electrons in only two dimensions (the simulation plane) for p-polarized laser wave whereas the electrons are heated in all three spatial directions in 3D. 
Theoretical analysis in Ref. \cite{Stark2017} demonstrated that this effect accounts for higher temperatures in 2D. 
The observation 3) can be explained by higher hole boring velocity in 3D calculations (which implies higher energy of protons accelerated by RPA) and reduced spread of hot electrons in 2D geometry \cite{Margarone2012} as well as higher temperature of hot electrons in 2D (which implies larger accelerating field generated by hot electrons leading to higher energy of protons accelerated by TNSA behind the target). 

\section{Influence of laser wave polarization}
Theoretical studies on RPA mostly propose higher stability of this mechanism when circularly polarized laser pulse is employed for the acceleration. 
The nonlinear force acting on the dense electron layer, where the laser wave is reflected, has a steady and oscillatory component in general case, whereas only steady component is present when the laser is circularly polarized. 
In the case of light sale acceleration (LS RPA), this should strongly reduce the heating of electrons which otherwise leads to unfavorable expansion of thin targets and limited time when the acceleration can occur \cite{Klimo2008,Robinson2008,Macchi2005}. 
However, the situation becomes more complicated in real 3D~case with limited size of the focused laser beam. 
The foil is bent by the laser beam intensity shape and hot electrons can be generated in the curved surface of the dense target. 
Moreover, Rayleigh-Taylor instability should develop \cite{Palmer2012,Lezhnin2015}. 
Similar effect can be observed in our case with a thicker target when we tested both polarizations (linearly p-polarized and circularly polarized pulses). 
In the initial stage of the interaction, hot electron heating is reduced for circularly polarized laser pulse. 
However, when the hole is bored in the target, the heating of hot electrons is relatively similar. 
For example, the estimated temperature of hot electrons in our simulations is about 34~MeV for linearly p-polarized pulse (LP) and about 32~MeV for circular polarization (CP) when the peak intensity interacts with the hydrogen target (140~fs from the beginning of laser-plasma interaction) whereas it is about 16~MeV vs. 2~MeV at earlier stage (60~fs from the beginning of the interaction). 

Due to strongly reduced heating of electrons in the earlier stage of the interaction for CP, proton acceleration in TNSA regime is also reduced here. 
Nevertheless, TNSA starts to work efficiently somewhat later and energies of protons accelerated by TNSA mechanism are similar at later stage for both polarizations. 
Timing and efficiency of TNSA and HB RPA for linearly and circularly polarized laser beams can be well illustrated by phase space of protons at the end of laser target interaction in 3D simulations in Fig. \ref{fig6} b), c). 
The position of the fastest TNSA protons originated from target rear side (located initially at $x = 15~\rm{\mu m}$) is about $x = 28~\rm{\mu m}$ for LP and about $x = 25~\rm{\mu m}$ for CP, although their maximum momentum is roughly the same (between 22 and 23~$m_ec$), which is a clear signature that TNSA was delayed for CP. 
The phase space around $x = 15~\rm{\mu m}$ also reveals that HB velocity is slightly enhanced for LP, which can be explained by the presence of the oscillatory component of ponderomotive force for LP. 

\begin{figure}[h]
\begin{center}
\includegraphics[width=0.45\textwidth]{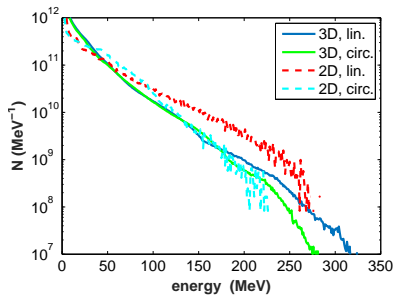}

\caption{\label{fig7} Energy spectra of accelerated protons for circularly and linearly polarized laser beams in 2D and 3D simulations at 260~fs from the beginning of laser-target interaction.}
\end{center}
\end{figure}

Proton energy spectra are drawn in Fig. \ref{fig7}, both from 2D and 3D~simulations. 
When the pulse is CP in 2D~simulations, HB RPA generates more protons in the mean energy range (around 70~MeV), although the number of protons in the high-energy range (from 100~MeV) and the maximum energy are lower compared with LP.  
These observations can be ascribed to the absence of the oscillatory component of ponderomotive force leading to a more stable acceleration and to reduced heating of hot electrons.
However, we have observed less pronounced difference between both polarization cases in more realistic 3D~simulations, where the heating of electrons and accelerating fields for TNSA are lower compared with 2D.

\section{Effect of laser pulse duration}
Up to now, we investigated the interaction of hydrogen target with laser pulse of duration exceeding 100~fs. 
However, multiPW laser beams with pulse duration of about 30~fs can be also expected in near future \cite{Danson2015}. 
Usually, the highest efficiency of proton acceleration with these ultrashort pulses can be observed by using ultrathin foils \cite{Ceccotti2007,Kim2016}. 
Here, we investigate a relative decrease in maximum energy and number of accelerated protons when the pulse becomes shorter and other parameters are kept constant. 

In our example, we set laser intensity to $3.7 \times 10^{21}~\rm{W/cm^2}$ and full laser pulse duration of $\sin^2$ temporal profile to 64~fs (similar to gaussian temporal profile of the pulse with FWHM equal to 30~fs).
Other parameters are set to the same values for initial 2D simulations introduced here (those with $25~\rm{\mu m}$ thick hydrogen target and various laser intensities). 

The results show the decrease of maximum energy of RPA protons from about 110~MeV for long pulse to about 70~MeV for short pulse, the number of high energy protons is reduced even more, by factor 5.3 in 2D simulations for short pulse. 
These observations are in agreement with our expectations assuming that laser beam burns the hole several times deeper for long pulse, thus the front of the laser pulse interacting with plasma (laser-plasma interface) can accelerate more protons. 
The laser beam is also focused inside the plasma when it burns the hole as shown in Fig. \ref{fig5}, which increases its intensity and explains the enhancement of maximum energy of RPA protons for long pulse, which is theoretically given by the dimensionless pulse amplitude $a_0$ of the focused beam.

\section{Conclusions}
We demonstrated high efficiency of proton acceleration from relatively thick solid hydrogen target (conversion efficiency almost 30~percent of laser pulse energy to protons accelerated in the forward direction in 2D and 3D simulations with laser and target parameters described above). 
Detailed analysis have shown that most of protons is initially accelerated in the target interior by HB RPA mechanism and they can be further accelerated behind the target by electrostatic field generated by hot electrons. 
Since the HB velocity increases with decreasing ion density and ion mass, hydrogen ribbon is the best candidate to demonstrate efficient HB RPA compared with other solid targets such as plastic or metal foils.

For laser intensities of the order of $10^{22}~\rm{W/cm^2}$ corresponding to laser power of several PWs and laser pulse length exceeding 100~fs at FWHM, the number of protons accelerated by HB~RPA and their maximum energies are much higher than the number of protons accelerated by TNSA from the rear side of the hydrogen ribbon. 
Even if a part of hydrogen target is evaporated and slightly expands before the arrival of the main laser pulse, thus the preplasma is formed, the effect of such preplasma is rather negligible on the final proton acceleration. 
Also, the polarization of laser wave does not affect significantly the final results for the assumed laser pulse length.

\subsection*{Acknowledgments}
This work has been mainly supported by the Czech Science Foundation, project No. 15-02964S. 
Fruitful discussions with Dr. D. Margarone and Prof. S. V. Bulanov from ELI-Beamlines project at IoP CAS are gratefully acknowledged. 
Computational resources were provided in the frame of IT4Innovations National Supercomputing Center supported by project LM2015070 and storage facilities were provided in the frame of CERIT Scientific Cloud supported by project LM2015085, both funded by Czech Ministry of Education, Youth and Sports. 
Partial support by the project LQ1606 funded by Czech Ministry of Education, Youth and Sports is also acknowledged. 

\bibliography{Hfoil}

\end{document}